# Optimized electrical control of a Si/SiGe spin qubit in the presence of an induced frequency shift

K. Takeda,[1] J. Yoneda,[1] T. Otsuka,[1, 2, 3] T. Nakajima,[1] M. R. Delbecq,[1, 4] G. Allison,[1] Y. Hoshi,[5] N. Usami,[6] K. M. Itoh,[7] S. Oda,[8] T. Kodera[9] and S. Tarucha[1, 10]

1. RIKEN, Center for Emergent Matter Science (CEMS), Wako-shi, Saitama, 351-0198, Japan
2. JST, PRESTO, 4-1-8 Honcho, Kawaguchi, Saitama 332-0012, Japan
3. Research Institute of Electrical Communication, Tohoku University, 2-1-1 Katahira, Aoba-ku, Sendai, 980-8577, Japan
4. Laboratoire Pierre Aigrain, Ecole Normale Supérieure-PSL Research University, CNRS, Université Pierre et Marie Curie-Sorbonne Universités, Université Paris Diderot-Sorbonne Paris Cité, 24 rue Lhomond, 75231 Paris Cedex 05, France
5. Advanced Research Laboratories, Tokyo City University, 8-15-1 Todoroki, Setagaya-ku, Tokyo 158-0082, Japan
6. Graduate School of Engineering, Nagoya University, Nagoya 464-8603, Japan
7. Department of Applied Physics and Physico-Informatics, Keio University, Hiyoshi, Yokohama 223-8522, Japan
8. Department of Physical Electronics and Quantum Nanoelectronics Research Center, Tokyo Institute of Technology, O-okayama, Meguro-ku, Tokyo 152-8552, Japan
9. Department of Electrical and Electronic Engineering, Tokyo Institute of Technology, O-okayama, Meguro-ku, Tokyo 152-8552, Japan
10. Department of Applied Physics, The University of Tokyo, Hongo, Bunkyo-ku, Tokyo, 113-8656, Japan

Correspondence: K. Takeda (kenta.takeda@riken.jp) or S. Tarucha (tarucha@ap.t.u-tokyo.ac.jp)

Abstract

Electron spins confined in quantum dots are an attractive system to realize high-fidelity qubits owing to their long coherence time. With the prolonged spin coherence time, however, the control fidelity can be limited by systematic errors rather than decoherence, making characterization and suppression of their influence crucial for further improvement. Here we report that the control fidelity of Si/SiGe spin qubits can be limited by the microwave-induced frequency shift of electric dipole spin resonance and it can be improved by optimization of control pulses. As we increase the control microwave amplitude, we observe a shift of the qubit resonance frequency, in addition to the increasing Rabi frequency. We reveal that this limits control fidelity with a conventional amplitude-modulated microwave pulse below 99.8%. In order to achieve a gate fidelity > 99.9%, we introduce a quadrature control method, and validate this approach experimentally by randomized benchmarking. Our finding facilitates realization of an ultra-high fidelity qubit with electron spins in quantum dots.

Introduction

Electron spins confined in semiconductor quantum dots provide an excellent platform for scalable solid-state quantum computing [1]. Quantum operations including single-spin rotation [2-4] and two-spin entanglement control [5-7] have been realized in the past. The control fidelities for single- [8-12] and two-qubit gates [13-16] have been largely improved by recent technical advancements in extending the spin coherence time. The single-qubit control fidelities have already reached the level close to or exceeding the threshold value required for implementing fault-tolerant logical qubits in the surface code structure [8, 10-16].

As the qubit performance improves, one needs to challenge the simplified view that relates spin qubit control fidelity solely to the ratio between the dephasing rate and the operation speed, since unitary errors such as pulse-induced effects can also be relevant. This problem has never been addressed, however, for quantum-dot qubits with a single electron spin 1/2 forming a natural two level system, in contrast to some other qubit systems where it is widely recognized (e.g. a.c. Stark shift and state leakage for transmons [17-19]). Such an approach may facilitate rapid single-qubit gates with fidelities high enough for fault-tolerant universal quantum operations [20], where multiple single-qubit gates are commonly involved for a two-qubit gate implementation. In addition, it is also important for precise qubit error metrology based on quantum tomography, which usually relies on single-qubit control for precise state preparation and measurement.

Here we report the observation and correction of microwave pulse induced systematic qubit errors in quantum-dot spin qubits. The spin qubit used in this work is defined in Si/SiGe quantum dots with a cobalt micro-magnet [21]. When the microwave burst is applied, in addition to the expected spin rotation, we observe an unexpected shift of the spin resonance frequency. While the frequency shift is typically an order of magnitude smaller than the Rabi frequency ($f_{\mathrm{Rabi}}$), it is much larger than the spin resonance linewidth and therefore causes a systematic error in the qubit rotation axis. This will limit the single-qubit control fidelity to 99.8 % according to our numerical simulations with realistic experimental parameters. To mitigate this problem and achieve high-fidelity, we introduce a quadrature microwave control which corrects the phase error of the qubit. The improvement of the qubit fidelity is experimentally confirmed by randomized benchmarking [22].

Results

The quantum dots used here are formed by locally depleting a two-dimensional electron gas in an undoped Si/SiGe heterostructure using lithographically defined electrostatic gates (Fig. 1a). We measure two devices, A and B, with a nominally identical structure except for the quantum well materials to characterize sample-to-sample dependence. The quantum well in device A has a natural isotopic composition [10] and for device B it consists of isotopically enriched silicon with approximately 800 ppm $^{29}$Si [12]. An on-chip cobalt micro-magnet induces the magnetic field gradient across the quantum dot [21]. A nearby sensor quantum dot coupled to a radio-frequency tank circuit allows rapid measurement of the quantum dot charge configuration [23]. All measurements were performed at an electron temperature of approximately 120 mK in a dilution refrigerator with an in-plane external magnetic field $B_{\mathrm{ext}}$. The spin state is read out in a single-shot manner using an energy-selective spin-to-charge conversion [24]. We use a quantum dot formed in the left (right) side of the device for device A (B). The expected lithographical dot position is shown as the blue (red) circle in Fig. 1a.

Figure 1b shows the pulse sequence for the spin control. First, a spin-down electron is prepared by applying gate voltages such that only the spin-down electron can tunnel into the dot. Next, the gate voltages are pulsed such that the electron confined in the dot is pushed deep in Coulomb blockade. Then, a microwave burst with a frequency of $f_{\mathrm{MW}}$ is applied to gate C to induce electric dipole spin resonance (EDSR). Finally, the gate voltages are pulsed back to the spin readout position where only a spin-up electron can tunnel out to the reservoir. When the microwave burst is applied to the gate, the electrons confined in the dot oscillate spatially in the slanting magnetic field induced by the micro-magnet, resulting in an effective oscillating magnetic field $B_{AC}$ perpendicular to the static magnetic field $B_0 = B_{ext} + B_z^{MM}$. At the condition where $hf_{MW} = g\mu_B B_0$ ($g$ is the electron $g$-factor and $\mu_B$ is the Bohr magneton), EDSR takes place. The inhomogeneous dephasing time of each qubit is estimated to be $T_2^* \sim 1.8$ μs for device A [10] and $T_2^* \sim 20$ μs for device B [12] from the Gaussian decay of the Ramsey fringe amplitude. In addition, device A has a Hahn echo decay time $T_2^H \sim 11$ μs (the associated measurement result is available in Supplementary Section 2) and device B has a Hahn echo decay time $T_2^H \sim 99$ μs [12].

The effect of strong EDSR microwave pulses can be readily observed in the microwave frequency dependence of the Rabi oscillations. Figure 1c shows the Rabi oscillation measured in device A with 3 different microwave amplitudes. $P_\uparrow$ is the spin-up probability obtained by averaging 500 to 1,000 single-shot measurement outcomes. The applied microwave burst has a rectangular envelope with an amplitude that is denoted by $A_{MW} = 0.3\sqrt{P(f_{MW})/P_0(f_{MW})}$, where $P(f_{MW})$ is the microwave power and $P_0(f_{MW})$ is the microwave power corresponding to $f_{Rabi} = 10$ MHz. The definition results in a normalized microwave amplitude of $A_{MW} = 0.3$ at $f_{Rabi} = 10$ MHz. For the smallest microwave amplitude ($A_{MW} = 0.1$), the resonance frequency is almost at the center of the image ($f_{MW} = 15.748$ GHz, indicated by the red arrows). However, when $A_{MW}$ is increased to 0.3, the center resonance frequency moves to higher frequencies. This frequency shift is further enhanced by increasing the microwave amplitude (~5 MHz frequency shift for $A_{MW} = 0.6$).

To quantify the resonance frequency shift $\Delta f$ more precisely, we perform a modified Ramsey interference measurement with an off-resonance microwave burst (Fig. 2a). It is worth noting that this measurement can also check whether the shift occurs only on resonance or not. During the waiting time $t_w$ between two resonant $X_{\pi/2}$ pulses, we apply an additional off-resonance microwave burst at a frequency of $f_{MW} = f_{res} - 180$ MHz, where $f_{res} = g\mu_B B_0/h$ is the bare qubit resonance frequency in the weak driving limit. When the qubit precession frequency shifts due to the off-resonance microwave burst, the oscillation period of the Ramsey fringe changes. Figure 2b shows the frequency shift $\Delta f$ for device B measured for various $A_{MW}$. Each data point is obtained by fitting the Ramsey oscillations using a sinusoidal function $P_\uparrow(t) = A\sin(2\pi\Delta f + \eta) + B$ with $A$, $B$, $\eta$ and $\Delta f$ as fitting parameters as shown in Fig. 2c (the data for device A is available in Supplementary Section 3). We find that an empirical power-law relation $\Delta f = aA_{\mathrm{MW}}^b$ fits well with the experimental data for both devices, however, the fitting parameters $a$ and $b$ are distinctively different between them. This may indicate that the frequency shift is related to some uncontrolled sample dependent parameters (e.g. local confinement potentials, defects etc.). We obtain the exponents $b = 1.39 \pm 0.02$ for device A (data shown in Fig. S3) and $b = 0.59 \pm 0.03$ for device B. Moreover, it is found that $\Delta f$ is positive ($a > 0$) for device A, while it is negative ($a < 0$) for device B.

An additional striking feature of the frequency shift is observed in the post microwave burst response. We find that, even after the microwave burst is turned off, the qubit resonance frequency shift remains and causes an additional qubit phase accumulation. To quantify this, the qubit phase accumulated after a microwave burst is extracted from a Hahn echo type measurement. Here we utilize a modified Hahn echo sequence which consists of two π/2 pulses, a π pulse and an additional 200 ns off-resonance microwave burst (Fig. 3a). The off-resonance microwave burst is interleaved in between the π pulse and the second π/2 pulse. The phase of the second π/2 pulse is modulated by $\phi$ to extract the echo phase $\theta(t_{\mathrm{d}})$. The post pulse delay time $t_{\mathrm{d}}$ indicates the time interval between the off-resonance microwave burst and the second π/2 pulse. The evolution time between the π/2 pulses and the π pulse is fixed to 20 μs to cancel out the unwanted phase fluctuation caused by quasi-static noise. Figure 3b shows the post pulse time dependence of the echo signal. Figure 3c shows the extracted echo phase evolution after the microwave burst application. For $A_{\mathrm{MW}} = 0$, the black solid line shows an average of the blue data points, while for $A_{\mathrm{MW}} = 0.15$, the black solid curve shows a fitting curve with an exponential function $\theta(t_{\mathrm{d}}) = C\exp(-t_{\mathrm{d}}/\tau) + D$ with $C$, $\tau$, and $D$ as fitting parameters, giving a characteristic decay time of $\tau = 6$ μs. For both cases, the offset at $t_{\mathrm{d}} = 0$ is mainly caused by the post-pulse phase accumulation due to the on-resonance pulses. From the measured qubit phase accumulation $\theta(t_{\mathrm{d}})$, the temporal post microwave burst frequency shift $\Delta f(t_{\mathrm{d}}) = (1/2\pi)(\mathrm{d}\theta(t_{\mathrm{d}})/\mathrm{d}t_{\mathrm{d}})$ can be obtained (Fig.3d). The green points show numerical derivative obtained from the data points in Fig. 3c. The black solid line shows an exponential fitting curve. Although the single exponential function fits the measured phase data well for $t_{\mathrm{d}} \geq 0.3$ μs, $\Delta f(t_{\mathrm{d}} = 0) \sim -80$ kHz derived from the single exponential dependence extrapolation does not match the value estimated from the fitting curve to the continuous-wave response derived from Fig. 2b ($\Delta f(t_{\mathrm{d}} = 0) \sim -320$ kHz with $A_{\mathrm{MW}} = 0.15$). We also note that the similar frequency shift as observed here was also measured in a different Si/SiGe spin qubit device with micro-magnet [16] and in a phosphorous donor electron spin qubit, albeit with values several orders of magnitude smaller [25].

There may be several physical origins for the frequency shift and among them we find that heating caused by the microwave burst may explain the exponential delayed response of the frequency shift (see Supplementary Sections 4 and 5). Since the thermal expansion is different between silicon and germanium, the increase of the lattice temperature can cause a change of the strain in the quantum well [26]. The strain caused by the metallic gate electrodes [27] may also be temperature dependent. In any case, the strain variation modifies the potential shape for the confined electron and the center quantum dot position. Because of the magnetic field gradient, the quantum dot position shift results in the local magnetic field or the resonance frequency shift. Since it takes some time to cool down the system to the base temperature after turning off the microwave burst, the frequency shift occurs during and even after the microwave burst application. However, this does not explain the discontinuous frequency shift between the continuous-wave response in Fig. 2b and the exponential decay in Fig. 3c because there should be no abrupt change in the system temperature before and after turning off the microwave pulse. Although the detailed physical mechanism will not affect the qubit fidelity optimization described in what follows, further investigation is needed to fully explain the observed frequency shift.

Now we turn to the qubit control fidelity. The observed resonance frequency shift affects the control fidelity because it is much larger than the fluctuation of resonance frequency for our device ($\sigma \sim 20.6$ kHz for device B). Therefore, here we discuss the qubit control optimization in the presence of such a microwave amplitude dependent frequency shift. The simplest way to cancel the frequency shift effect may be to keep the microwave amplitude always constant by applying off-resonance microwave even when the qubit is idle [16]. In this way, the qubit frequency shift during the control stage is kept constant and we can choose the shifted qubit resonance frequency as the rotating frame frequency. However, this method causes too much additional heating of the device which may be harmful for the qubit control because we need a relatively large microwave power to realize the qubit rotation faster than the dephasing time. In addition, due to the limited bandwidth of the microwave modulation circuit, creation of the smooth shaped pulse is difficult for this type of control including abrupt frequency switching.

We therefore investigate a way to cancel out the unwanted qubit phase accumulation by quadrature microwave control [17, 19, 28]. The technique was originally proposed for cancelling the microwave induced frequency shift (a.c. Stark shift) and the state leakage of transmon qubits. Because spin qubits generally have a well-defined two-level system and the state leakage is negligible, the quadrature control can be used to just correct the microwave induced frequency shifts. In this case, in contrast to the transmon qubit case where the single quadrature parameter has to be set to an optimal point to balance the compensation of two infidelity sources, one quadrature parameter can be used to fully compensate the influence of the frequency shift. To calculate the single-qubit time evolution, here we consider the rotating frame Hamiltonian of the system written as follows:

$$H(t) = -\frac{\hbar}{2}(X(t)\sigma_x + Y(t)\sigma_y + Z(t)\sigma_z),\tag{1}$$

where $X(t)$ and $Y(t)$ are the EDSR microwave control amplitudes, $Z(t)$ is the frequency shift caused by the XY control, and $\hbar$ is the reduced Planck's constant. The rotating frame frequency and $f_{MW}$ are set at the qubit resonance frequency during the free evolution with $X(t) = Y(t) = 0$. Here we consider the pulse optimization for a Gaussian π/2 rotation $X(t) = A_X \exp(-t^2/2\sigma^2)$ and the quadrature derivative control $Y(t) = \alpha_{\pi/2}\sigma(dX(t)/dt)$ truncated at $\pm 2\sigma$. $A_X$ is the microwave control amplitude normalized with the ideal π/2 control amplitude $A_{\pi/2} = \pi/(\sigma \int_{-2}^{2} \exp(-t^2/2)\,dt)$. Note that the quadrature coefficient $\alpha$ has to be adjusted independently for π and π/2 pulses. The microwave induced frequency shift is calculated from the power-law relation $Z(t) = a(X(t)^2 + Y(t)^2)^{b/2}$ ($t \in [-2\sigma, 2\sigma]$), i.e. it is assumed to be dominated by the instantaneous response and the slowly changing part is ignored. The partial optimization still works reasonably well to mitigate the qubit control errors because the slow delayed response is several times smaller than the fast response.

Figure 4a shows a plot of the averaged qubit control fidelity $\bar{F}$ of $X_{\pi/2}$ gate calculated using the equation $\bar{F}(U, \mathcal{E}) = 1/2 + (1/12)\sum_{j=x,y,z} \text{Tr}(U\sigma_j U^\dagger \mathcal{E}\sigma_j)$, where $U = \exp(i\pi\sigma_x/4)$ is the ideal process matrix and $\mathcal{E}$ is the actual quantum operation [29]. Here we plot $\bar{F}$ for the gate clock frequency $t_{\pi/2}^{-1} = 1/4\sigma$ ranging from 1 to 20 MHz, which is a reasonable operation range for device B. In this qubit operation range, $\bar{F}$ is limited to approximately 99.8 % because of the unwanted phase accumulation due to the frequency shift. In Fig. 4b, we calculate $\bar{F}$ at $t_{\pi/2}^{-1} = 20$ MHz (corresponds to $f_{Rabi} = 5$ MHz for rectangular microwave burst) as a function of π/2 quadrature coefficient $\alpha_{\pi/2}$. The model predicts a gate fidelity higher than 99.999 % with an optimized parameter set at $A_X = 1.00$ and $\alpha_{\pi/2} = -0.173$. (The graphical Bloch sphere representation of the qubit evolution is depicted in Fig. S6.) We experimentally confirm the effectiveness of the quadrature control using an interleaved randomized benchmarking technique (Fig. 4c). Only device B is used for this measurement as the influence of the frequency shift is too subtle to observe experimentally in device A. The $X_{\pi/2}$ interleaved randomized benchmarking is used to characterize the fidelity of $X_{\pi/2}$ gate and $f_{MW}$ is set to the free evolution frequency calibrated by the Ramsey fringe. Figures 4d and 4e show the $X_{\pi/2}$ interleaved randomized benchmarking sequence fidelity $F$ at a fixed number of Clifford gates, $m = 122$, measured for various values of $\alpha_{\pi/2}$ and $A_X$. The sequence fidelity is defined as $F = P_\uparrow^{|\uparrow\rangle} - P_\uparrow^{|\downarrow\rangle}$, where $P_\uparrow^{|\uparrow\rangle}(P_\uparrow^{|\downarrow\rangle})$ is the measured spin-up probability for the sequence designed to obtain $|\uparrow\rangle(|\downarrow\rangle)$ as an ideal final state. To clarify the parameter dependence of $\alpha_{\pi/2}$ and $A_X$, the other parameters (microwave frequency and amplitude, $\alpha$ for other Clifford gates) are adjusted to maximize the sequence fidelity. We find that the sequence fidelity is maximized at $\alpha_{\pi/2} = -0.18$, which is in reasonable agreement with the value derived from the theory. The small deviation may come from the post pulse effect. From a separate measurement using the same device and the quadrature control, we obtain a single gate fidelity as high as 99.93 % [12] and this is well above the upper limit given by the microwave burst induced frequency shift.

## Discussion

We have reported the shift of resonance frequency of electron spin qubits in Si/SiGe quantum dots with increasing applied microwave burst amplitude and quadrature control method to cancel out the qubit control error cause by the frequency shift. Although part of the observed frequency shift may be explained by the effect of heating, the overall physical origin remains unknown and full characterization needs further investigation. Nevertheless, for the purpose of practical optimization of quadrature compensation pulse presented in this work, the Ramsey-based measurement of the amplitude dependence described in Fig. 2 is sufficient. We anticipate that the full understanding of the frequency shift mechanism will allow for further optimizations beyond what is presented in this work, such as the prediction of the frequency shift from the device parameters and the minimization of the frequency shift itself by the device design.

## Methods

In both devices, the quantum dot is formed by locally depleting the two-dimensional electron gas in an undoped Si/SiGe heterostructure. A 250 nm thick cobalt micro-magnet is deposited on top of the accumulation gate to induce a stray magnetic field across the quantum dot. The sample is cooled down using a dilution refrigerator to a base electron temperature of approximately 120 mK (unless otherwise noted) which is estimated from the transport linewidth. Further details about the devices and the measurement setup are described in Supplementary information, Refs. 10 (device A), and 12 (device B).

For both devices, the valley splitting is confirmed by magneto-spectroscopy measurement to be larger than the Zeeman splitting. Therefore, the physics in this work is mainly described by a conventional single-valley picture, although there may be a small fraction of the population in the excited valley state due to initialization errors.

### Data availability
The data that support the findings of this study are available from the corresponding author upon reasonable request.

### Acknowledgements

This is a post-peer-review, pre-copyedit version of an article published in npj Quantum Information. The final authenticated version is available online at: https://doi.org/10.1038/s41534-018-0105-z. We thank the Microwave Research Group in Caltech for technical support. This work was supported financially by Core Research for Evolutional Science and Technology (CREST), Japan Science and Technology Agency (JST) (JPMJCR15N2 and JPMJCR1675) and the ImPACT Program of Council for Science, Technology and Innovation (Cabinet Office, Government of Japan). K.T. acknowledges support from JSPS KAKENHI grant number JP17K14078. J.Y., T.O. and T.N. acknowledge support from RIKEN Incentive Research Projects. T.O. acknowledges support from Precursory Research for Embryonic Science and Technology (PRESTO) (JPMJPR16N3), JSPS KAKENHI grant numbers JP16H00817 and JP17H05187, Advanced Technology Institute Research Grant, the Murata Science Foundation Research Grant, Izumi Science and Technology Foundation Research Grant, TEPCO Memorial Foundation Research Grant, The Thermal and Electric Energy Technology Foundation Research Grant, The Telecommunications Advancement Foundation Research Grant, Futaba Electronics Memorial Foundation Research Grant and Foundation for Promotion of Material Science and Technology of Japan (MST) Foundation Research Grant. K.M.I. acknowledges support from JSPS KAKENHI grant number JP26220602 and JSPS Core-to-Core Program. T.K. acknowledges support from JSPS KAKENHI grant numbers JP26709023 and JP16F16806. S.T. acknowledges support from JSPS KAKENHI grant numbers JP26220710 and JP16H02204.

**Competing interests**

The authors declare that they have no competing interests.

**Author contribution**

K.T. and J.Y. performed the measurements and analyzed the data. K.T., and T.O. fabricated the samples. Y. H., N. U., and K. M. I. supplied the isotopically enriched Si/SiGe heterostructure. K.T. wrote the article with inputs from the rest of the authors. M.R.D., G.A., T.N., T.K., and S.O. contributed to the sample fabrication, measurement, and data analysis. S.T. supervised the project.

Figures and tables

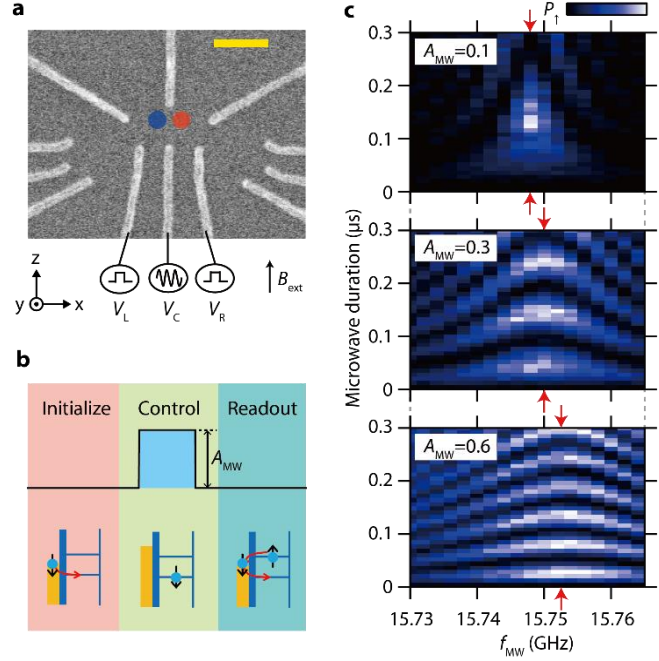

**Figure 1.** Device structure and Rabi oscillation frequency shift. **a** Scanning electron microscope image of the device. The scale bar represents 200 nm. The gate electrode geometry is nominally identical for both devices A and B. Three of the gate electrodes (R, L, and C) are connected to the 50 ohm coaxial lines. The blue (red) circle shows the estimated position of the quantum dot for device A (B). **b** Pulse sequence used for the Rabi oscillation measurement. The initialization and readout are done at the same gate voltage condition where only the spin-down electron can tunnel into the dot. The compensation stage to make the pulse d.c. voltage offset to zero (used only for device A) is omitted for simplicity. **c** Rabi oscillation measured with different microwave amplitudes at $B_{\text{ext}} = 0.51$ T (device A). The red arrows show the center resonance frequency positions. As $A_{\text{MW}}$ is increased, in addition to the increase of $f_{\text{Rabi}}$, the center resonance frequency increases as well.

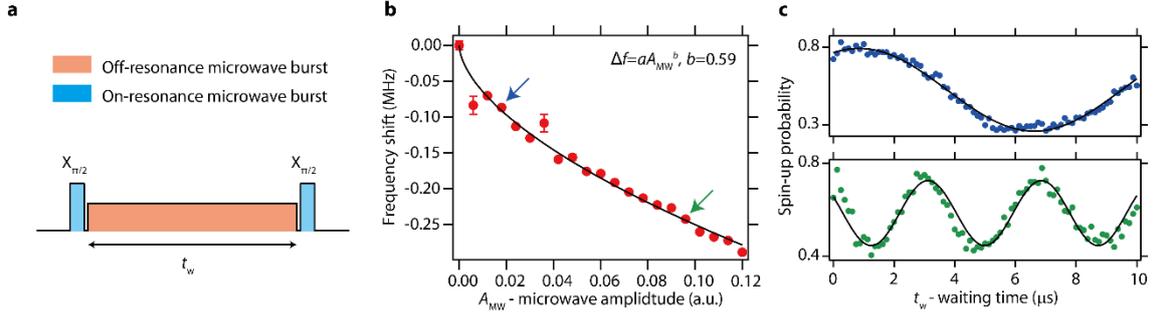

**Figure 2**. Resonance frequency shift measurements (device B). **a** Schematic showing the modified Ramsey sequence. During the waiting time $t_\text{w}$, an off-resonance microwave burst with a rectangular envelope is applied to observe the microwave induced frequency shift. **b** Resonance frequency shift $\Delta f$ measured as a function of the off-resonance microwave amplitude $A_\text{MW}$. The red points show the experimental data and the black solid line shows a power-law fitting $\Delta f = a A_\text{MW}^b$ with $b = 0.59$. **c** Ramsey fringe oscillations measured under the conditions indicated by the arrows in Fig. 2b. The black solid lines show sinusoidal fitting curves.

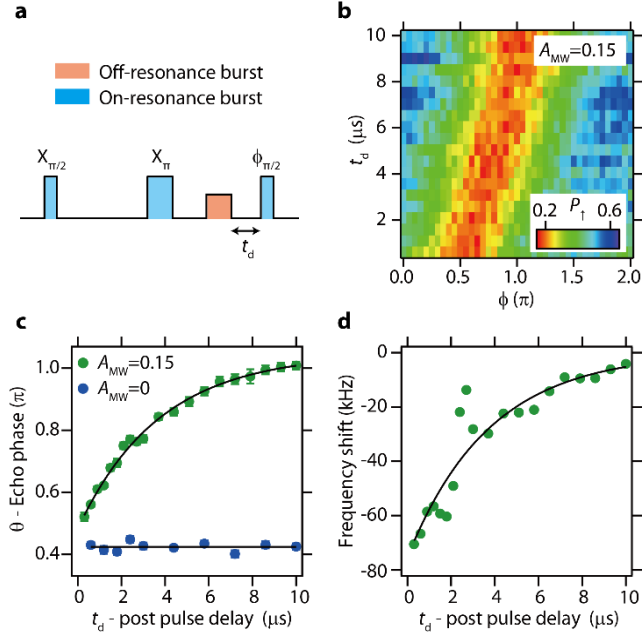

**Figure 3**. Post pulse frequency shift measurement (device B). **a** Schematic showing the modified Hahn echo sequence used to obtain the post microwave burst response. The interval between each $\pi/2$ pulse and the $\pi$ pulse is fixed at 20 μs. **b** Measured echo signal shift as a function of $t_d$ at $A_{MW} = 0.15$. **c** Extracted echo phase shift $\theta$ after turning off the microwave burst. The circles show the data obtained by fitting the echo signal with a sinusoidal function $P_\uparrow(\phi) = -E\cos(\phi + \theta(t_d)) + F$ with $E(>0)$, $F$, and $\theta(t_d)$ as fitting parameters. The error bars represent one standard deviation of uncertainty. The black solid lines show fitting curves. **d** Transient frequency shift derived from the echo phase accumulation at $A_{MW} = 0.15$. The black solid line shows a derivative of the exponential fitting curve $\Delta f(t_d) = (1/2\pi)(d\theta(t_d)/dt_d)$ in Fig 3c.

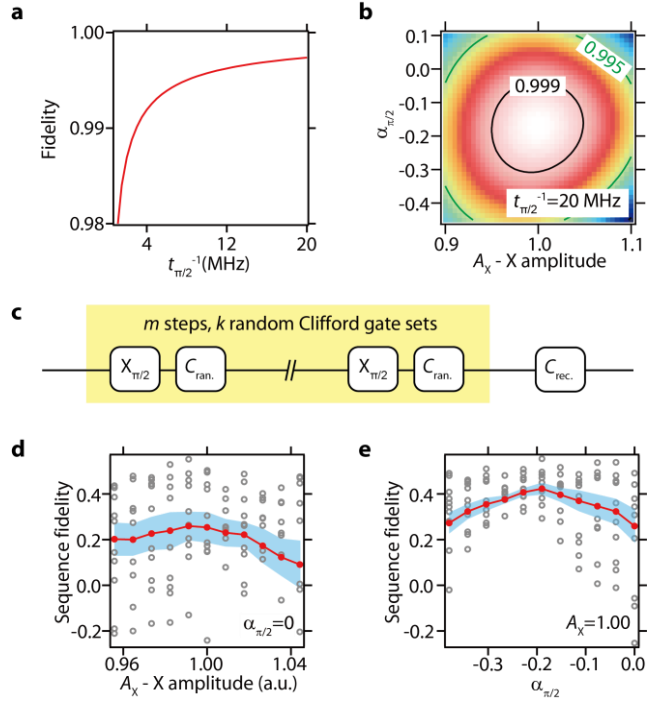

**Figure 4.** Qubit fidelity analysis and optimization using a quadrature microwave control technique. **a** Calculated qubit operation clock $t_{\pi/2}^{-1}$ dependence of the averaged qubit fidelity $\overline{F}$ of $X_{\pi/2}$ gate. **b** Average qubit fidelity $\overline{F}$ as a function of the control amplitude and the quadrature control coefficient $\alpha_{\pi/2}$. The gate time is set at $t_{\pi/2} = 50$ ns. **c** Schematic showing the pulse sequence for the randomized benchmarking measurement. In between the randomly chosen Clifford gates, an $X_{\pi/2}$ gate is interleaved to characterize its fidelity. **d** Interleaved randomized benchmarking fidelity for $X_{\pi/2}$ gate measured as a function of X control amplitude $A_X$. $A_X$ is directly proportional to the microwave voltage amplitude. The number of random Clifford gates is fixed at $m = 122$ and $k = 32$ gate sets are used for the measurements. The grey scattered points show the sequence fidelity for each random Clifford gate set and the red points show the sequence fidelity averaged over all 32 random gate sets. The light blue band shows standard error of the mean at each $A_X$. **e** Interleaved randomized benchmarking fidelity for $X_{\pi/2}$ gate measured as a function of the quadrature coefficient $\alpha_{\pi/2}$. The number of random Clifford gates is fixed at $m = 122$. The grey scattered points show the sequence fidelity for each random Clifford gate set and the red points show the sequence fidelity averaged over all 32 random gate sets. The light blue band shows standard error of the mean at each $\alpha_{\pi/2}$.

# Supplementary information for "Optimized electrical control of a Si/SiGe spin qubit in the presence of an induced frequency shift"

K. Takeda,[1] J. Yoneda,[1] T. Otsuka,[1, 2, 3] T. Nakajima,[1] M. R. Delbecq,[1, 4] G. Allison,[1] Y. Hoshi,[5] N. Usami,[6] K. M. Itoh,[7] S. Oda,[8] T. Kodera[9] and S. Tarucha[1, 10]

1. RIKEN, Center for Emergent Matter Science (CEMS), Wako-shi, Saitama, 351-0198, Japan
2. JST, PRESTO, 4-1-8 Honcho, Kawaguchi, Saitama 332-0012, Japan
3. Research Institute of Electrical Communication, Tohoku University, 2-1-1 Katahira, Aoba-ku, Sendai, 980-8577, Japan
4. Laboratoire Pierre Aigrain, Ecole Normale Supérieure-PSL Research University, CNRS, Université Pierre et Marie Curie-Sorbonne Universités, Université Paris Diderot-Sorbonne Paris Cité, 24 rue Lhomond, 75231 Paris Cedex 05, France
5. Advanced Research Laboratories, Tokyo City University, 8-15-1 Todoroki, Setagaya-ku, Tokyo 158-0082, Japan
6. Graduate School of Engineering, Nagoya University, Nagoya 464-8603, Japan
7. Department of Applied Physics and Physico-Informatics, Keio University, Hiyoshi, Yokohama 223-8522, Japan
8. Department of Physical Electronics and Quantum Nanoelectronics Research Center, Tokyo Institute of Technology, O-okayama, Meguro-ku, Tokyo 152-8552, Japan
9. Department of Electrical and Electronic Engineering, Tokyo Institute of Technology, O-okayama, Meguro-ku, Tokyo 152-8552, Japan
10. Department of Applied Physics, The University of Tokyo, Hongo, Bunkyo-ku, Tokyo, 113-8656, Japan

## S1 Device structure and micro-magnet field simulation

Both devices A and B are fabricated on top of undoped Si/SiGe heterostructures. The details of the Si/SiGe heterostructures are available in Refs. 10 (device A) and 12 (device B). Figure S1a shows the layer stack of our Si/SiGe quantum dot device. The surface of the heterostructure is covered by a 10 nm thick $Al_2O_3$ insulator formed by atomic layer deposition. The ohmic contacts are fabricated by phosphorus ion implantation. The quantum dot confinement gates and the accumulation gate are formed by electron-beam lithography and metal deposition. The accumulation gate and depletion gate electrodes are separated from each other by another 50 nm thick $Al_2O_3$ insulator layer. A 250 nm thick cobalt micro-magnet is deposited on top of the accumulation gate to induce a stray magnetic field around the quantum dot.

The micro-magnet magnetic field simulation shown here is performed by using the Mathematica package Radia [30]. The following micro-magnet field simulation results are based on the geometry of device A (60 nm thick SiGe spacer layer), but device B (40 nm thick SiGe spacer layer) has a very similar geometry and therefore the conclusions are quantitatively applicable to both devices. Figures S1a and S1b show the device layer stack and the micro-magnet geometry. The blue point in Fig. S1b shows the ideal quantum dot position, however, due to the limited alignment precision in electron-beam lithography and the non-ideality of the gate voltage confinement, we speculate that there may be about ± 50 nm uncertainty in the quantum dot position. Figs. S1c and S1d show the calculated slanting field $dB_y^{MM}/dz$ distribution. Around the expected position of the quantum dot ($x = -0.05$ μm and $z = 0$ μm), $dB_y^{MM}/dz \sim 0.7$ T/um is obtained.

## S2 Echo measurement for device A

Figure S2a shows the Hahn echo measurement sequence. Figure S2b shows the echo signal as a function of evolution time $t_{\text{evolve}}$. By fitting the normalized echo decay $C(t_{\text{evolve}})$ using $C(t_{\text{evolve}}) = \exp(-(t_{\text{evolve}}/T_2^H)^\alpha)$ with $T_2^H$ and an exponent α as fitting parameters, we obtain $T_2^H = 11$ μs and α = 1.2.

## S3 Frequency shift measurement for device A

The pulse sequence used for the measurement is depicted in Fig. 2a in the main text. First, a $X_{\pi/2}$ pulse (−7.4 MHz detuned from the center resonance frequency) is applied to rotate the spin state to the equator of the Bloch sphere. Next the frequency shift is induced by an off-resonance microwave burst (−67.4 MHz detuned from the center resonance frequency) with a duration of $t_w$. Finally the accumulated qubit phase is projected to the z-axis by the second $X_{\pi/2}$ pulse (−7.4 MHz detuned). The −7.4 MHz frequency offset is intentionally added to observe a finite frequency oscillation at $A_{MW} = 0$. Figure S3a shows the measured oscillations of the Ramsey fringe as a function of off-resonance microwave amplitude $A_{MW}$. Figure S3b shows the frequency shift extracted by fitting the observed Ramsey data as a function of $A_{MW}$. The red points show the data and the black solid line shows an empirical fitting curve with $\Delta f = a A_{MW}^b$ with $a$ and $b$ as the fitting parameters. The exponential function fits well to the measured data, however, as mentioned in the main text, it is found that the fitting constants have strong sample-to-sample variation.

## S4 Possible reasons for the frequency shift

1. Bloch-Siegert shift

The most conventional reason for the microwave induced qubit resonance frequency shift may be the Bloch-Siegert shift, which is well-known as a special case of a.c Stark shift for strongly driven two-level systems [31]. The frequency shift we observed is obviously inconsistent with this mechanism because in this case the frequency shift should follow the quadratic relation $\Delta f = f_{Rabi}^2 / E_Z$ where $E_Z = g\mu_B B_0$ is the Zeeman splitting. In addition, the value of the shift is quantitatively too small (device A: $\Delta f \sim 6$ kHz for $f_{Rabi}$=10 MHz and $E_Z = 16$ GHz, device B: $\Delta f \sim 2$ kHz for $f_{Rabi}$=5 MHz and $E_Z = 18$ GHz).

2. Rectification due to the anharmonicity of the confinement

Another possible reason is the quantum dot motion rectification due to the anharmonicity of the confinement potential. Figure S1e shows the calculated in-plane stray magnetic field ($B_z$) distribution in the two-dimensional electron gas (2DEG) plane. Although the micro-magnet is designed to be robust against a relatively large misalignment [33], the robustness only stands for the slanting field $dB_y^{MM}/dz$, not for the Zeeman splitting or the magnetic field $B_z$. As shown in Fig. S1f, misalignment of the micro-magnet along the z-direction can cause the relatively large variation of the in-plane longitudinal field gradient $dB_z^{MM}/dz$. For ±50 nm misalignment of the micro-magnet position, $dB_z^{MM}/dz$ can be as large as 0.3 T/μm. In addition, the direction of electric field by gate C at the quantum dot position is not perfectly parallel to the z-axis due to the device geometry and therefore the transverse field gradient $dB_z^{MM}/dx$ (Fig. S1g) may also contribute to the gate voltage dependence of $B_z$. Here, using the value of $dB_z^{MM}/dz = 0.3$ T/μm, we numerically simulate the rectification effect with a one-dimensional model assuming quartic anharmonic potential $U(z) = \sum_{k=2}^{4} a_k z^k$. Although it is difficult to directly measure the coefficients $a_3$, $a_4$ experimentally, we can estimate these parameters by the Rabi frequency saturation caused by the anharmonicity. In both devices A and B, the anharmonicity suppresses $f_{Rabi}$ to ~19 MHz at $A_{MW} = 0.6$, which is approximately 1 MHz smaller than the value expected from the linear relation observed at lower microwave amplitudes [10, 12]. Figure S4a shows simulated Rabi frequency suppression $\Delta f_{Rabi}(a_3, a_4) = f_{Rabi}(0,0) - f_{Rabi}(a_3, a_4)$ and Fig. S4b shows the exponent of the resonance frequency shift obtained by fitting the Rabi frequency deviation with a power-law relation $\Delta f = a(A_{MW})^b$. In the proper parameter range where $\Delta f_{Rabi}(a_3, a_4)$ is close to the experimental value 1 MHz (the black region in Fig. S4a), we obtain $b \sim 2$, which results in a quadratic shift (Fig. S4b). There is a clear discrepancy between the simulated and the measured exponents $b$ ($b = 0.59 \pm 0.03$ for device B and $b = 1.39 \pm 0.02$ for device A). Therefore, we rule out the rectification effect from the possible reasons for the frequency shift.

3. External microwave setup

The frequency shift can also occur if there is a frequency shift of the microwave signal. Such an unintentional microwave frequency change might be caused by a large output parameter change of the microwave signal generator (Keysight E8267D is used for all measurements). However, when we change $A_{MW}$ in the experiment, rather than changing the output condition (power) of the signal generator, we change the output amplitude of an arbitrary waveform generator used for I/Q modulation. This modulation scheme causes hardly any change of the frequency shift of the microwave signal. In addition, we monitor the output microwave signal (after passing all active components) using a spectrum analyzer and there is no noticeable shift of the frequency when the I/Q modulation amplitude or the source power is changed. The microwave circuit contains some additional passive components (attenuators, cables etc.), but those will not affect the frequency of microwave signal.

As for the delayed response of the frequency shift, the reflection of the microwave due to impedance mismatch is a possible cause. The microwave reflection can cause delayed transient microwave output with the time scale depending on the length of the reflection path. In our measurement setup, the dominant source for the microwave reflection seems to be the room-temperature modulation circuit as the I/Q mixer and the microwave amplifier(s) have much poorer VSWRs as compared to those for the other components. At the output of the microwave modulation circuit, we measured transient signal with a time scale < 10 ns using a real-time oscilloscope (Keysight DSOX92004Q with a sampling rate of 80 GSa/s). Although it may slightly affect the experimental results, the time scale is much shorter than we observed in Fig. 3c in the main text (a characteristic decay time $\tau \sim 6$ μs) and cannot explain the experimental result.

4. Heating due to microwave burst

Here, we consider a Hamiltonian with harmonic confinement as follows:

$$H = \frac{\hat{p}_u^2 + \hat{p}_v^2}{2m^*} + \frac{1}{2\hbar^2} E_{\text{orb}}^2 m^* (\hat{u}^2 + \hat{v}^2) - e\vec{E} \cdot \begin{pmatrix} \hat{u} \\ \hat{v} \end{pmatrix}, \tag{1}$$

where $m^*$ is the transverse effective mass of electron in strained silicon, $e$ is the elementary charge, $\hbar$ is the reduced Planck's constant, $\hat{p}_u$ and $\hat{p}_v$ are the momentum operators, $\hat{u}$ and $\hat{v}$ are the position operators, $E_{\text{orb}}$ is the orbital spacing of the quantum dot, and $\vec{E}$ is the in-plane electric field. In what follows, we set the electric field as $\vec{E} = (E, 0)^{\text{T}}$ for simplicity, but it does not affect the conclusion due to the symmetry of the Hamiltonian (note that we have assumed a symmetric in-plane confinement). From simple mathematics, the dot position offset $u_0$ caused by $\vec{E}$ can be derived as follows:

$$u_0 = \frac{\hbar^2 eE}{m^* E_{\text{orb}}^2}. \tag{2}$$

According to Ref. 32, the potential change caused by the temperature change can be dealt as the change of the effective mass. In such a case, the induced dot position shift $\delta u_0$ for a small change of effective mass $\delta m^*$ can be written as follows:

$$\delta u_0 = |u_0(m^* + \delta m^*) - u_0(m^*)| \sim \frac{\hbar^2 eE}{m^* E_{\text{orb}}^2} \left(\frac{\delta m^*}{m^*}\right). \tag{3}$$

Then, the position shift results in a frequency shift when combined with the micro-magnet field gradient as follows:

$$\delta f = \gamma_e \frac{dB_z^{\text{MM}}}{du} \delta u_0 \sim \gamma_e \left(\frac{dB_z^{\text{MM}}}{du}\right) \frac{\hbar^2 eE}{m^* E_{\text{orb}}^2} \left(\frac{\delta m^*}{m^*}\right), \tag{4}$$

where $\gamma_e = 28$ GHz/T is the gyromagnetic ratio for electron spin in silicon. By using a parameter set $dB_z^{\text{MM}}/du = 0.3$ T/μm, $E = 0.1$ MV/m, $m^* = 0.19 m_e$ ($m_e$ is the electron rest mass), and $E_{\text{orb}} = 0.5$ meV, in addition to the referred effective mass change caused by the strain effect $\delta m^*/m_e = 2 \times 10^{-6} \times \delta T$ $K^{-1}$, where $\delta T$ is the temperature change [32], we can obtain an estimate for the frequency shift,

$$|\delta f| \sim 14 \times \delta T \text{ kHz/K}. \tag{5}$$

According to this equation, the measured electron temperature increase of a few hundred mK under the microwave burst will cause a few tens of kHz frequency shift, which is not too far from the measured results in Fig. 3d (we assume thermal equilibrium between the lattice and electrons). However, we stress that this result just shows a rough estimation of the frequency shift value and is not enough to deal with the detailed properties like the device dependent exponent and sign. Simulations including the local electrostatics and strain of the devices will be needed for further investigation.

### S5 Measurement of electron temperature under microwave burst

For both devices examined in this study, the microwave burst is applied to a gate (C gate in Fig. 1), which has an electrically open end. Ideally, the voltage applied on the open end circuit is reflected back to the source and there should be no power consumption at the sample end. However, the high frequency signals (15-20 GHz) used for the EDSR measurements in this work will be easily dissipated. For instance, the stray capacitances to the neighboring conductors (the surrounding two-dimensional electron gas and metallic gates) can result in finite a.c. current flowing through the device. To confirm this, we perform a measurement of electron temperature under the microwave burst. Device A is used for the measurement and the microwave frequency is 20 MHz detuned from the spin resonance frequency. Fig. S5a shows the dot-to-reservoir transitions under several different microwave excitation amplitudes. As can be seen in the data, the line width is broadened with the increased microwave amplitude. From the Fermi-Dirac fitting curve, we can extract $T_\text{e}$ at each $A_\text{MW}$. Fig. S5b shows $T_\text{e}$ extracted for a wider range of $A_\text{MW}$. The measurement is limited to a relatively low-power range ($A_\text{MW} = 0.012$ corresponds to $f_\text{Rabi} = 400$ kHz) because the charge sensor sensitivity rapidly decreases at the higher microwave amplitudes and the reliable estimation of the line width becomes difficult. We find that the line width increases linearly as a function of the microwave amplitude. From this measurement, although at higher temperatures some cooling mechanisms will suppress the linear temperature increase, we roughly estimate that the electron temperature increases by a few Kelvin when a microwave burst for EDSR in the MHz range is applied. We note that this measurement is done at a decreased base electron temperature with a modified setup ($T_\text{e} \sim 35$ mK in this measurement whereas $T_\text{e} \sim 120$ mK for the others).

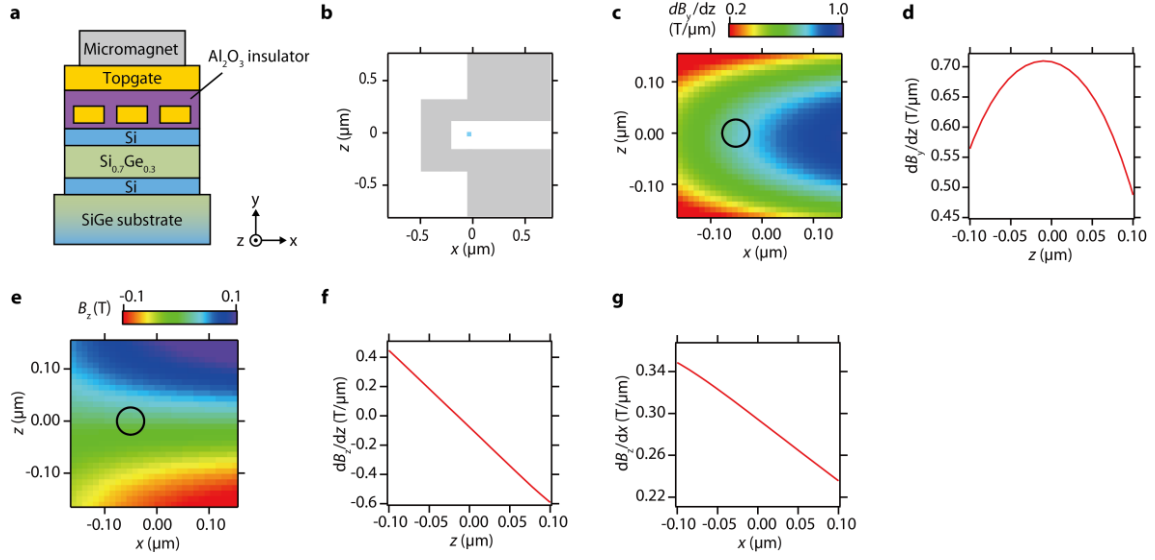

**Figure S1.** Micro-magnet design and simulation results. **a** Schematic layer sequence of the device structure. The external magnetic field is applied along the positive z-direction. **b** Schematic of the micro-magnet design. The grey area shows the micro-magnet pattern and the quantum dot location is represented by the blue box. **c** Simulated out-of-plane slanting magnetic field $dB_y^{MM}/dz$ as a function of the positions in the 2DEG plane $x$ and $z$. The black circle shows the expected quantum dot position. The designed quantum dot position is $x = -0.05$ μm and $z = 0$ μm. **d** Line cut of the out-of-plane slanting field $dB_y^{MM}/dz$ along the z-axis at the dot position $x = -0.05$ μm. **e** Simulated in plane stray magnetic field $B_z^{MM}$ as a function of the positions in the 2DEG plane $x$ and $z$. The black circle shows the expected quantum dot position. The designed quantum dot position is $x = -0.05$ μm and $z = 0$ μm. **f** Line cut of $dB_z^{MM}/dz$ along the z-axis at the dot position $x = -0.05$ μm. **g** Line cut of $dB_z^{MM}/dx$ along the x-axis at $z = 0$ μm.

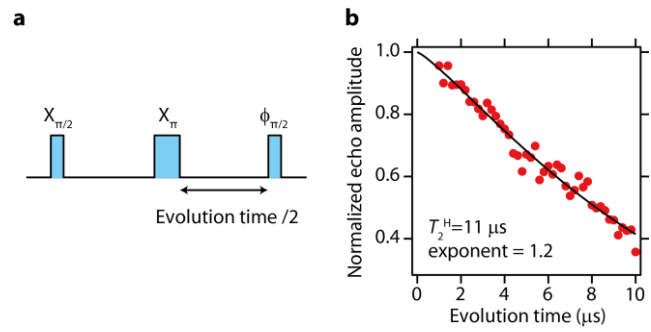

**Figure S2.** Spin echo measurement for device A. **a** Schematic of the echo measurement pulse sequence. **b** Measured echo data. The red circles show the data points and the black solid line shows a fitting curve.

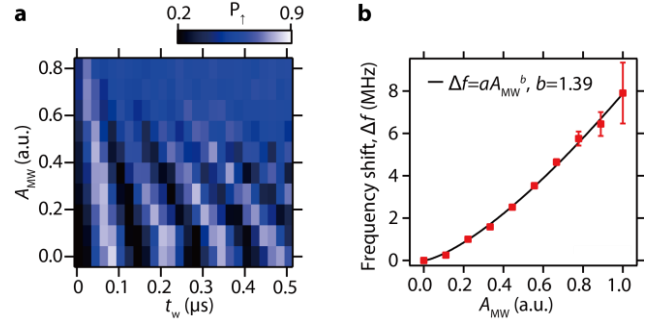

**Figure S3.** Modified Ramsey measurement result for device A. **a** Ramsey fringe measurement result at $B_{\text{ext}} = 0.5045$ T and $f_{\text{MW}} = 15.600$ GHz. **b** Measured resonance frequency shift $\Delta f$ as a function of the off-resonance microwave amplitude $A_{\text{MW}}$. The red points show the experimental data and the black solid line shows a power-law fitting $\Delta f = a A_{\text{MW}}^b$ with $b = 1.39$.

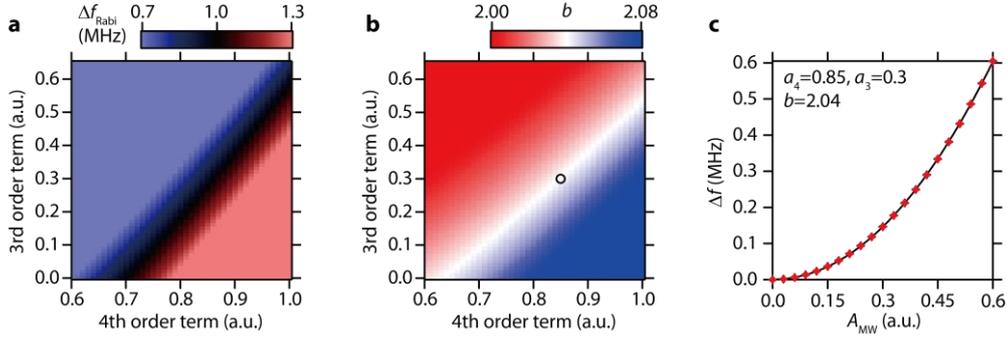

**Figure S4**. Simulation results for the effects of anharmonicity. **a** Rabi frequency suppression $\Delta f_{\text{Rabi}}$ calculated as a function of the third and fourth order coefficients. The microwave amplitude is set to $A_{\text{MW}} = 0.6$, which corresponds to $f_{\text{Rabi}} = 20$ MHz if a linear relation is assumed. The black coloured area shows the conditions where the calculated values are close to the experimental value $\Delta f_{\text{Rabi}} \sim 1$ MHz. **b** Calculated exponents $b$ as a function of the third and fourth order term strengths. $b \sim 2$ is obtained for all parameter value used in the simulation. **c** A typical microwave amplitude dependence of the frequency shift with parameters $a_3 = 0.3$ and $a_4 = 0.85$ (the black circle in Fig. S4b).

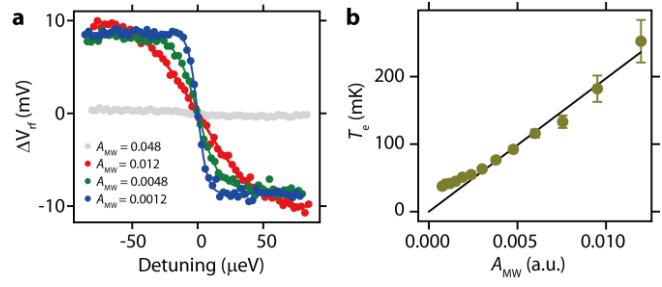

**Figure S5**. Measurement of electron temperature under microwave excitation. **a** Dot-to-reservoir transition measured at four different microwave amplitudes. The circles show the measured data. The solid lines show fitting curves with a Fermi-Dirac function $\Delta V_{\rm rf} = \frac{V}{2}\tanh\left(\frac{\varepsilon}{2k_{\rm B}T_{\rm e}}\right)$ with $V$ and $T_{\rm e}$ as fitting parameters. A linear background is subtracted from each of the data sets and the fitting curves. **b** Electron temperature as a function of the microwave amplitude. The circles show the electron temperatures extracted from the dot-to-reservoir transition line widths and the linear fitting curve $T_{\rm e} \propto A_{\rm MW}$ is obtained by using the six data points from the largest $A_{\rm MW}$.

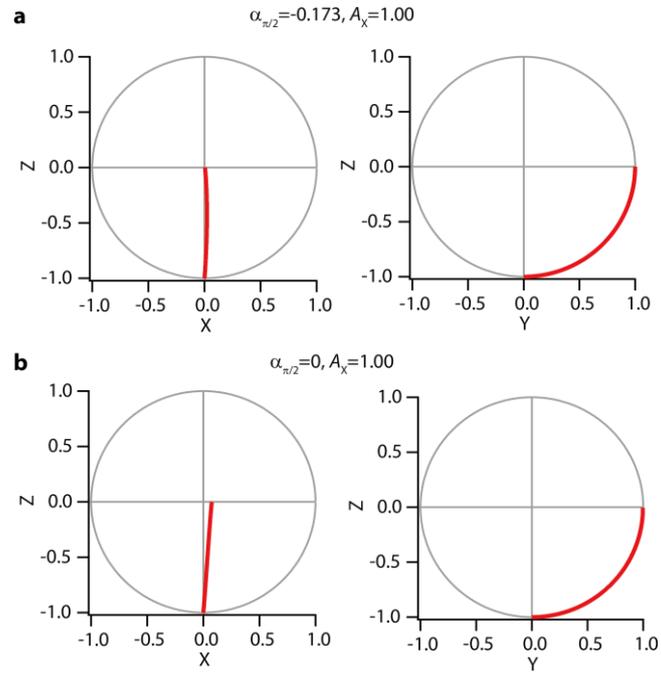

**Figure S6.** Bloch sphere representations of qubit trajectory for $X_{\pi/2}$ pulse for an initial state of spin down. **a** Optimized control with quadrature control $\alpha_{\pi/2} = -0.173$ and $A_X = 1.00$. **b** Standard control without quadrature control $\alpha_{\pi/2} = 0$ and $A_X = 1.00$.